\documentclass[pdflatex,photonics,article,moreauthors]{Definitions/mdpi} 
\usepackage{comment} 
\usepackage{outlines} 
\usepackage{graphicx}
\usepackage{caption}
\usepackage{glossaries}
\usepackage{xcolor}         
\usepackage{multirow}
\usepackage{float}

\firstpage{1} 
\pubvolume{1}
\issuenum{1}
\articlenumber{0}
\pubyear{2024}
\copyrightyear{2024}
\datereceived{12.03.2024} 
\daterevised{15.03.2024} 
\dateaccepted{29.05.2024} 
\datepublished{} 
\hreflink{https://doi.org/} 

\Title{Machine Learning in Short-Reach Optical Systems: A Comprehensive Survey}

\TitleCitation{Machine Learning in Short-Reach Optical Systems: A Comprehensive Survey}

\Author{
Chen Shao $^{1}$\orcidA{},
Elias Giacoumidis $^{2}$\orcidC{},
Syed Moktacim Billah $^{3}$, 
Shi Li $^{3}$\orcidD{},
Jialei Li $^{3}$,
Prashasti Sahu $^{4}$, 
André~Richter $^{3}$\orcidF{},
Michael Faerber $^{1}$\orcidG{},
Tobias Kaefer $^{1}$\orcidH{}
}

\AuthorNames{Chen Shao, Syed Moktacim Billah, Elias Giacoumidis, Jialei Li, Prashasti Sahu, André~Richter, Michael Faerber, Tobias Kaefer}

\AuthorCitation{C.Shao.; M. B Syed.; }

\address{%
$^{1}$ \quad Karlsruhe Institute of Technology (KIT), Germany; \{chen.shao2, jia.li9, michael.faerber, tobias.kaefer\}@kit.edu\\
$^{2}$ \quad Karlsruhe Institute of Technology (KIT), Germany; syedsakibgp@gmail.com\\
$^{3}$ \quad VPIphotonics GmbH, Hallerstraße 6, 10587 Berlin, Germany; \{Shi.Li, elias.giacoumidis, andre.richter\}@vpiphotonics.com \\
$^{4}$ \quad Technical University of Chemnitz, Str. der Nationen 62, 09111 Chemnitz, Germany; prashastisahu13@gmail.com} 

\abstract{
Recently, extensive research has been conducted to explore the utilization of machine learning (ML) algorithms in various direct-detected and (self)-coherent short-reach communication applications. These applications encompass a wide range of tasks, including bandwidth request prediction, signal quality monitoring, fault detection, traffic prediction, and digital signal processing (DSP)-based equalization. As a versatile approach, ML demonstrates the ability to address stochastic phenomena in optical systems networks where deterministic methods may fall short. However, when it comes to DSP equalization algorithms such as feed-forward/decision feedback equalizers (FFE/DFE) and Volterra-based nonlinear equalizers, their performance improvements are often marginal, and their complexity is prohibitively high, especially in cost-sensitive short-reach communications scenarios such as passive optical networks (PONs). Time-series ML models offer distinct advantages over frequency domain models in specific contexts. They excel in capturing temporal dependencies, handling irregular or nonlinear patterns effectively, and accommodating variable time intervals. Within this survey, we outline the application of ML techniques in short-reach communications, specifically emphasizing their utilization in high-bandwidth demanding PONs. We introduce a novel taxonomy for time-series methods employed in ML signal processing, providing a structured classification framework. Our taxonomy categorizes current time series methods into four distinct groups: traditional methods, Fourier convolution-based methods, transformer-based models, and time-series convolutional networks. Finally, we highlight prospective research directions within this rapidly evolving field and outline specific solutions to mitigate the complexity associated with hardware implementations. We aim to pave the way for more practical and efficient deployment of ML approaches in short-reach optical communication systems by addressing complexity concerns.
}


\begin{document}

\section{Introduction}
\label{sec:intro}

Short-reach optical transmission systems have gained substantial attraction owing to their remarkable attributes of high bandwidth and low latency \cite{ref-short_reach_communication}. In the evolving landscape of communication technologies, short-reach optical communication has emerged as an essential domain, driven by the increasing demand for high-speed data transfer in applications such as inter-data centers  \cite{ref-data_center}, access/local area networks and industrial automation \cite{xie2022machine}. This increasing demand requires efficient, low-latency communication systems tailored to short-reach scenarios, typically up to 100 km. While long-haul, optical communication has been immersive in data transmission, its applicability encounters challenges when adapting to the constraints of shorter distances. This is mainly due to physical and technical limitations that prevent its seamless integration into existing networking environments characterized by the need for energy-efficient, and cost-effective data transmission over limited distances. Passive optical networks (PONs) utilize passive optical splitters and combiners, which are less expensive than the active components required in traditional point-to-point fiber networks. This makes PONs a cost-effective fiber-optic solution.

Since PONs rely on passive optical splitters, they inherently introduce power losses, limiting the overall power budget and the number of users that can be supported on a single PON. In addition, effects caused by the fiber, such as chromatic dispersion (CD) and nonlinearity can limit the PON-reach \cite{ref-optical_comm_challanges}, especially when intensity-modulated and direct-detected (IMDD) high baud-rate signals are considered \cite{ref-performance_factor}. 

Ongoing research endeavors are dedicated to advancing optical detection schemes to overcome these limitations and increase the signal bit-rate in both short-reach and long-haul optical communication networks \cite{ref-comm_schemes}. For instance, the regeneration of 
coherent optical systems in the last decade has been a major breakthrough, as they have gone beyond just using intensity-only modulation \cite{ref-next_gen_comm}. Coherent systems employ external modulators to employ complex baseband signals to the optical field. The optical coherent receiver, equipped with phase diversity, linearly recovers signals and compensates for fiber impairments through digital signal processing (DSP) \cite{ref-CD_compensation}. Coherent technology enables the transmission of advanced modulation formats and polarization multiplexing to increase the signal bit-rate significantly. Additionally, coherent optical systems enable dense wavelength division multiplexing (DWDM) and super-channels, which are pushing long-distance optical networks into the multi-terabit per second capacity range \cite{ref-comm_capacity}. 

Except for traditional homodyne-coherent technology, coherent communication strategies include diverse techniques, such as phase detection through heterodyne detection. While this approach has its merits \cite{ref-IMDD}, a notably favored incoherent approach such as IMDD is practically preferred due to its inherent simplicity and cost-effectiveness in short-reach communications \cite{ref-IMDD2,ref-IMDD3}. 

In contrast to coherent transmission, IMDD operates by encoding information into the intensity of the optical signal, with the modulation signal being real-valued and positive \cite{ref-IMDD3}. The implementation of IMDD eliminates the need for complex optical components and local oscillators, reducing hardware complexity. Additionally, IMDD systems are less susceptible to phase noise and polarization-related issues, making them robust and practical for scenarios where cost efficiency and simplicity are paramount \cite{ref-IMDD3}. 
Furthermore, practical considerations like operation and safety can limit the highest and average values of the modulated signal in IMDD systems. These restrictions give IMDD systems specific characteristics in how they function \cite{ref-IMDD4}. Various models, such as the Poisson channel, square-root Gaussian channel, and Gaussian channel with input-dependent noise, among others, exist to rapidly assess and characterize IMDD systems \cite{ref-IMDD5,ref-IMDD6,ref-IMDD7}. In contrast to conventional methodologies that depend on analog components and processing \cite{ref-conventional_method}, IMDD can potentially integrate machine learning (ML) algorithms at the receiver DSP if required \cite{ref-cost_effective}, providing a flexible and adaptable solution for enhancing the transmission performance. According to \cite{ref-hardware_complexity}, the combination of ML and DSP techniques allows IMDD systems to dynamically adapt and optimize signal parameters. This addresses impairments and variations in real-time without needing complex hardware adjustments. This approach represents a significant benefit, as it not only reduces the costs associated with complex hardware setups in short-reach systems, but also highlights the effectiveness of intelligent signal processing \cite{ref-advantage_of_ML,ref-cost_effective, ref-ML}.

In this survey, we examine the significant progress made in short-distance optical communications research over the past decade. First, we summarize several key research areas (Section 2). Afterwards, we focus on the equalization problem, introducing benchmark DSP methods (Section 3) and ML algorithms (Section 4). Then, we categorize recent sequence models in the ML field (Section 5), dividing them into convolution-based, transformer-based, and Fourier-based neural networks. We explore the advantages, disadvantages, and complexities of each method in addressing the equalization problem. In the final section, we provide an overview of the model compression field, outlining two approaches to compress models. We see these approaches as potential solutions for addressing hardware complexity concerns.

The primary contribution of this survey is to summarize the existing research on ML implementations for short-reach optical communications across a range of applications. Specifically, our contributions are the following: 
\begin{enumerate}
    \item We review existing Deep Learning (DL) models, providing a comprehensive understanding of their principles, characteristics, and hypothesis classes. This facilitates an in-depth exploration for researchers seeking supervised neural-network-based ML models suitable for their specific applications.  
    \item  We highlight the features and complexities of these models, elucidating recent developments in the field of DL. This information is valuable for researchers interested in delving deeper into research and staying abreast of current advancements. 
    \item  We discuss the current limitations and research gaps in the ongoing development of DL, addressing the challenges posed by these factors in real-world applications. Furthermore, we provide constructive insights regarding the selection of models and potential future directions.
    \item  Given the challenge of high hardware complexity, we introduce model compression as a potential solution from the DL field. 
We present existing works that employ this approach within the optical communication field, aiming to inspire more researchers to pursue research in this domain. 
\end{enumerate}

\section{Applications in Short-Reach Systems}
\label{sec:Applications in short reach communication system}

After systematically organizing recent literature in the past few years, we have categorized ML-based research for short-reach optical systems into four classes based on application tasks: Bandwidth Request and Prediction, Subcarrier Allocation, Equalization, and Fault Detection. We clarify the physical and mathematical aspects of their respective tasks, enumerate several recent works, and provide a summary of current advancements.

\textbf{Bandwidth Request and Prediction:} It aims to leverage network information to predict future bandwidth availability and enable its utilization by related applications. In mathematical terms, the real-time bandwidth forecast at a specific time (t) involves estimating the available bandwidth that will be accessible in the immediate future (t + $\tau$) \cite{qi2012dynamic}. One proposed method, known as predictive-dynamic bandwidth allocation (P-DBA), utilizes this concept to predict high-priority traffic during waiting periods, resulting in reduced latency and packet loss rates within a Gigabit PON (GPON) \cite{qi2012dynamic}. Another approach demonstrated in \cite{sarigiannidis2016dama} leverages the k-nearest neighbor algorithm to predict additional bandwidth requirements for each optical network unit (ONU) in a PON. This adaptive learning-based approach dynamically adjusts the k value based on real-time traffic conditions, showcasing the adaptability of ML in optimizing bandwidth allocation \cite{sarigiannidis2016dama}. Artificial neural networks (ANNs) have also shown promise in achieving flexible bandwidth allocations across various application scenarios, particularly emphasizing low-latency objectives \cite{ruan2018machine, yi2019machine}. For example, feed-forward-based ANNs, explored in \cite{mikaeil2018traffic}, are utilized to predict packet arrivals in time-division multiple access (TDMA) ONUs, effectively reducing additional DBA processing delays \cite{mikaeil2018traffic}. Furthermore, Xgboost\cite{xgboost} is employed to predict bandwidth requests for ONUs in Ethernet PON (EPON), optimizing bandwidth utilization across polling periods. This study introduced a dynamic wavelength and bandwidth assignment scheme for time and WDM (TWDM) PONs, incorporating regression techniques for efficient resource allocation \cite{ye2018recurrent}. Recent studies show that ML approaches are versatile in addressing challenges related to predicting and managing bandwidth needs. This paves the way for developing more adaptive and efficient short-reach optical communication systems in the near future \cite{qi2012dynamic, sarigiannidis2016dama, ruan2018machine, yi2019machine, mikaeil2018traffic, ye2018recurrent}.

\textbf{Subcarrier Allocation:} The optimization of bandwidth allocation for enhanced spectral efficiency has led to increased interest in subcarrier allocation for PONs. This approach involves mathematically formulating the allocation problem as an integer linear programming (ILP) task, which includes tasks such as optimizing wavelength configurations, assigning subcarriers to transmitters, and minimizing lost traffic and energy costs. To address this challenge, deep reinforcement learning has emerged as a promising technique that enables dynamic subcarrier sharing among ONUs, facilitating efficient DBA. At the medium access control (MAC) layer, the dynamic subcarrier allocation (DSA) algorithm schedules ONU upstream transmissions by considering instantaneous bandwidth requirements and existing traffic conditions \cite{kanonakis2012physical} . This showcases the adaptability of ML in resource scheduling. Several studies focus on algorithm-level cost reduction and two-dimensional resource scheduling for orthogonal frequency-division multiplexing (OFDM)-PONs including \cite{kanonakis2012physical, lim2017dynamic, bi2013joint}. These DSA algorithms address challenges related to latency, throughput, and energy efficiency, highlighting the versatility of ML in enhancing subcarrier allocation strategies \cite{zhu2022dynamic}. Moreover, the integration of traffic prediction technology and fair-aware DSA algorithms, as proposed in \cite{zhu2022dynamic} and \cite{senoo2017fairness}, further enhances the performance of subcarrier allocation in short-reach optical communication systems. These advancements improve the efficiency and adaptability of subcarrier allocation by applying ML methodologies \cite{nakayama2022real}.

\textbf{Power Budget Limitations:} The electric power budgeting issue is about predicting future energy consumption using historical data on power usage and related environmental factors like weather, user behavior, and equipment efficiency. The goal is to forecast power consumption for upcoming time periods. However, the development of large-scale, systematic ML models for this task is limited by the lack of publicly available datasets. Recent research has provided a basic process for constructing the necessary data and has also presented baseline ML models as a starting point. Specifically, the data construction process involves compiling and organizing relevant datasets, including time-series power consumption data, weather information, occupancy patterns, and equipment performance metrics. This standardized data can then be used to develop and test ML models for power consumption forecasting. For instance, the recent work in \cite{power} has introduced baseline ML models that demonstrate the feasibility of using these techniques to predict future power consumption, despite the constraints posed by the scarcity of publicly accessible datasets.

\textbf{Equalization:} The objective of this task is to minimize fiber-induced distortions by employing post-processing techniques that compensate for linear effects, such as CD. Mathematically, the equalizer optimizes the function $f(x)$ to ensure that the equalized output sequence $y$ closely approximates the input signal. Performance evaluation primarily relies on the bit error ratio (BER). In PON systems, using shallow-based DL models for post-equalizers has shown potential in addressing nonlinear distortions for both IMDD and coherent signals. This is especially useful in scenarios with modulator nonlinearities or high-launched optical power to meet tight power budgets \cite{ref-next_gen_comm}. As the fiber-induced nonlinear effects are increasing in the latter case, in single-channel coherent PONs, this results in self-phase modulation (SPM). In multi-channel PONs, the increased nonlinear effects result in cross-phase modulation (XPM) and four-wave mixing (FWM). In IMDD PONs, low-complexity artificial neural network (ANN)-based equalizers have demonstrated performance comparable to Volterra-based equalizers in pulse amplitude modulation with 4 levels (PAM4) systems \cite{chen2022real}. While post-equalization techniques have proven effective, the computational complexity at the ONU receiver is a challenge. To address this, strategies for centralized pre-equalization at the transmitter side have been proposed. Examples include memory polynomial-based pre-equalizers \cite{chen2022real} and trained neural network-based pre-equalizers \cite{xue2021soa}. These methods enhance equalization effectiveness while keeping the ONU receiver simple. 

\textbf{Fault Detection:} Short-reach optical communication systems, including PONs, are susceptible to failures such as fiber cuts, equipment failures, power outages, natural disasters, and ONU transceiver malfunctions \cite{ss10207489}. Service disruptions can result in significant financial losses for service providers. Identifying faulty ONUs presents challenges, especially when nearly equidistant branch terminations lead to overlapping reflections, making it difficult to pinpoint the exact defective branch \cite{ss10207489}. Conventional monitoring approaches become less reliable as PON systems grow in complexity. Recent advancements in ML-enabled proactive fault monitoring offer promising solutions to ensure stable network operation. ML-based fault prediction algorithms utilize past network fault data to discover underlying patterns and similarities. By doing so, these algorithms enhance the detection of optical network problems and facilitate proactive repairs, thereby preventing potential issues from occurring. Several research papers propose using ML algorithms for monitoring management in optical networks. Notably, technologies like random forest and ANN algorithms have been employed to continuously monitor the BER, predict network component failures, and assess fault severity \cite{vela2018soft}. Wang et al. \cite{wang2017failure} introduced a hybrid approach combining double exponential smoothing and support vector machines for equipment failure prediction in software-defined metropolitan area networks. Bayesian network-based models have also been developed for diagnosing PON faults \cite{wang2017failure}.

\section{DSP for Signal Equalization in Communication Systems}
\label{sec:DSP for Signal Equalization in Communication Systems}
In this section, we provide an overview of conventional signal equalization techniques, ranging from basic zero-forcing equalization to more advanced approaches such as feed-forward equalizers (FFE), decision-feedback equalizers (DFE), Viterbi and Volterra equalizers, as well as adaptive equalizers. We discuss the advantages and limitations of these techniques, comparing the performance of ML models. Table \ref{tab:tab1} provides the complexity analysis for each method.

\textbf{Zero Forcing}: It is a linear equalizer (LE) derived by minimizing inter-symbol interference (ISI). A study in \cite{ref-Thuft} has established the analytical foundation for optimal zero-forcing and minimum mean squared error (MSE) equalization in channels with additive white noise and specified frequency response. The study demonstrates that an optimal LE can be implemented as a cascade of filters, with taps spaced at symbol intervals. However, when the channel effect exhibits a deep frequency response “valleys,”  equalization will yield poor performance due to noise enhancement.

\textbf{Feed Forward Equalizer:} FFE \cite{munagala2017novel} mitigates ISI in communication channels by processing the received signal forwardly without feedback. Its simplicity makes it suitable for systems where feedback is unstable or challenging for implementation.

\textbf{Decision Feedback Equalizers:} Due to the noise enhancement, DFE is designed to reduce ISI by subtracting already-known symbols. In this way, ISI from already detected symbols is eliminated. Adaptation of the forward and feedback filters of DFEs follow the same pattern as for LEs \cite{williamson1992block}. The disadvantage is that it could potentially lead to accumulated errors from feeding back incorrect detection decisions 

\textbf{Viterbi equalizer:}  The Viterbi equalizer seeks to estimate the most likely sequence of transmitted symbols, given the received sequence. By constructing a trellis diagram where nodes represent possible transmitted symbols and transitions denote potential channel transitions, the Viterbi algorithm dynamically optimizes path metrics to identify the most probable sequence. This process involves state transition probabilities and precise calculations to mitigate the impact of channel impairments. Mathematically, the Viterbi equalizer applies the Viterbi algorithm, which belongs to the dynamic programming algorithm for finding the most likely sequence of hidden states in a hidden Markov model. The time complexity of the Viterbi equalizer is determined by the Viterbi algorithm, which depends on the length of the input sequence and the number of states, making it $O(T \cdot N^2)$, where $T$ denotes the length of the sequence and $N$ refers to the number of hidden states \cite{vierto}.

\textbf{Volterra equalizer}: This is a nonlinear equalizer used in optical communication systems to compensate for nonlinear distortions introduced by the fiber channel \cite{volterra}. In PAM4 systems, severe ISI can be introduced due to the imperfect bandwidth of optical and electrical components. The main bandwidth bottleneck in IMDD systems comes from the transmitter side, as the achievable bandwidth of receiver-side devices is typically twice as high as the bandwidth of transmitter-side devices. In such scenarios, the Volterra equalizer can be effectively employed to address both a potential nonlinearity from the transmitter and the bandwidth limitations of the optical components. The higher-order Volterra kernels can model the frequency-dependent distortion and nonlinear effects caused by the limited transmitter bandwidth and nonlinear devices, such as Mach-Zehnder modulators. The Volterra equalizer is based on the Volterra series expansion, which allows for the modeling of nonlinear systems. The key idea is to use a set of nonlinear filters, known as Volterra kernels, to capture the nonlinear characteristics of the channel. The structure of a Volterra equalizer consists of multiple stages, each representing a different order of nonlinearity. The first stage corresponds to the linear equalizer, which performs initial equalization to address linear distortions. Subsequent stages of the Volterra equalizer capture and compensate for higher-order nonlinear distortions. These stages involve nonlinear filters that take multiple past symbols as inputs and produce outputs based on their interaction. The number of stages and the complexity of the Volterra equalizer depend on the specific system requirements and the level of nonlinear distortions present. The coefficients of the Volterra kernels are typically adapted or optimized using algorithms such as the least mean squares (LMS) or recursive least squares (RLS) algorithms. These algorithms iteratively adjust the coefficients based on the error between the equalized signal and the desired signal, aiming to minimize the distortion and improve the overall system performance.

\textbf{Adaptive filtering:} Adaptive filtering \cite{malik2011adaptive} is used in communication systems where channel characteristics vary over time. The mathematical interpretation involves using an adaptive algorithm that iteratively modifies the filter parameters to minimize the error signal between the desired output and the actual output, enabling the filter to adapt to changing input conditions. The actual convergence time and the total time complexity over multiple iterations depend on the convergence behavior of the specific algorithm and its sensitivity to the input data. Assuming t taps, the total time complexity for updating all coefficients is \textit{O}(\textit{t}). FFE and DFE are regarded as adaptive filtering versions designed explicitly for short-reach communications.

\begin{table}[H]
  \centering
  \tiny
  \caption{Complexity Analysis for DFE, FFE, LE, Adaptive Filtering and Viterbi algorithms. $t$ refers to the number of the taps. $N$ in Viterbi denotes the number of the hidden states.}
  \label{tab:tab1}
\footnotesize   
\begin{tabular}{c|cccll}
\toprule
Models &  DFE & FFE & LE   &Adaptive Filtering & Viterbi\\
\midrule
Train          &  $O(t)$ &  $O(t)$ & $O(t)$ &$O(t)$  & $O(t \cdot N^2)$\\
Inference  &  $O(t)$ &  $O(t)$ & $O(t)$ &$O(t)$ & $O(t  \cdot N^2)$\\
\bottomrule
\end{tabular}
\end{table}

\section{Traditional Sequential ML Methods}
\label{sec:Traditional Sequential ML Methods}
With the increasing demand for higher data transmission rates and the limitations of traditional prediction methods reaching their practical limits in terms of accuracy, the need for algorithms with high precision, reliability, and low complexity has become urgent. In this section, we introduce new DL-based models to address this challenge. We overview relevant research studies, providing a chronological exploration of key sequential models, namely, recurrent neural networks (RNN), long short-term memory (LSTM), gated recurrent unit (GRU), and convolutional neural networks (CNN). The key architectural parts of DL models are explained, with clear examples showing how they work, how they are used, and how complex they are.

In 2018, Karanov et al. \cite{karanov2018end} introduced an end-to-end deep neural network system for optical communications, encompassing the entire chain of a transmitter, receiver, and channel model. This research showed that transceiver optimization can be done in a complete, end-to-end way. Owing to the sequential structure of communication systems, sequential models, including LSTM networks \cite{graves2012long}, RNNs \cite{pascanu2013construct}, and GRUs \cite{jing2019gated} have been extensively employed. They are considered as baseline algorithms in order to generate more advanced and efficient algorithms.

\textbf{RNN:} Originally designed for machine translation in natural language processing, this model is based on the Markov assumption about the hidden state and output sequence: the output sequence depends only on the current potential state $h_t$. The potential state depends on the previous moment's latent $h_{t-1}$ and input variables $x_{t-1}$ rather than on the historical data ${x_{(t-1, \dots, 0)}, h_{(t-1, \dots, 0)}}$. Renowned for their adaptability in handling variable-length sequences and preserving state information across elements, these models find valuable applications in diverse communication fields \cite{xie2022machine}. In recent work, they have shown promising results in equalization compared to benchmark methods based on Volterra and Viterbi equalizers in two-dimensional eight-level PAM (2D-PAM8) links \cite{ref-xinqin}.

Despite its great equalization performance, this model suffers from exploding gradient issues caused by the direct gradient flow of multiple layers \cite{ref-rnngra}. In such networks, the backpropagation of the gradient is done by accumulating the gradient matrix. This can cause the gradient to grow exponentially if the eigenvalues of the gradient matrix are greater than 1, making the training process very difficult to converge. Conversely, when the eigenvalues of the gradient matrix are less than 1, the gradient will decrease over time until it vanishes completely, causing the parameters to stop updating \cite{RNNDiff}.

\textbf{LSTM:} The LSTM architecture can assist in overcoming this issue by extending the hidden state to a cell state, which is built using a gating mechanism. This mechanism has input, forget, and output gates that help control the flow of information \cite{ref-lstm}. LSTM models have additional internal states beyond just the hidden state. This allows them to learn a weight matrix that can better preserve useful information in the hidden state. The input gate decides what new information from the current input to be stored in the cell state. The forget gate decides what memories from the previous cell state to keep or discard. The output gate controls what information gets passed to the next cell state. This gating mechanism provides the ability to effectively hold onto relevant details from long sequences, while filtering out irrelevant information. This makes it easier to learn dependencies between distant parts of the input. As a result, LSTMs have been widely used in short-range communication tasks that require capturing complex long-term relationships in the data \cite{ling2022channel}.

\textbf{GRU:} This architecture simplifies the gating mechanism used in LSTM models. It has an update gate and a reset gate, instead of the three gates in LSTM \cite{dey2017gate}. The update gate determines what relevant information to retain from the previous state and the current input. The reset gate controls what data to discard. It is useful in scenarios where the temporal dependencies and relationships between adjacent symbols in a sequence are important. For example, in short-range communication systems, GRUs can help mitigate signal distortions caused by CD and nonlinearities \cite{deligiannidis2021performance}. Recent research in 120 Gb/s coherent 64-quadrature amplitude modulated optical systems for transmission at 375 km, has shown that using a bi-directional GRU as a nonlinear equalizer can help improve the quality factor (Q-factor) beyond the 8.52 dB limit (8.52 dB estimated from $Q = 20\log_{10}(\sqrt{2}\text{erfc}^{-1}(2\text{BER}))$) ~\cite{liu2021bi}, typically required for hard-decision forward error correction (HD-FEC). 

\textbf{CNN:} CNNs are not technically considered sequential models. However, they are widely used across many different domains. This is because of its important advantages, such as high parameter efficiency, weight-sharing mechanism, and plug-and-play characteristics \cite{wu2019cnn}. CNNs use a convolutional kernel to scan the input signal in a specific dimension, capturing temporal features that are important for the task at hand. This convolutional layer is typically followed by a pooling layer and a nonlinear activation function. The pooling layer reduces redundancy, while the activation function introduces nonlinearity. The convolutional kernel is designed to extract features that closely match the input signal, afterwards, backpropagation is used to optimize the weights of the network. This allows the CNN to learn and enhance the features that are most relevant for the target task. The weights in the network's weight matrix are updated through backpropagation to amplify the important features needed for effective performance on the given ML problem.  Furthermore, it has been observed that using multiple layers of small convolutional kernels is often more efficient than using large kernels. This approach, known as the inception architecture, was first introduced in the GoogleNet model \cite{GoogleNet}. Two commonly used blocks in CNN are the inception module and the inception reduction module, which extract temporal dependencies of different scales by employing a concatenation of $1 \times 1$, $3 \times 3$, and $5 \times 5$ convolutional kernels. In addition, It also uses a special type of convolutional kernel with a size of $1 \times 1$. This $1 \times 1$ convolution serves a unique purpose - it helps to reduce the number of feature map channels or dimensions. It is commonly used between two regular convolution layers or at the output layer of the network. 

\textbf{Summary:} 
In this section, we have introduced the most common building blocks used in ML models for short-reach optical communication systems. These fundamental components are still widely used in current approaches. To summarize the complexity of the models discussed earlier, we have provided a table (Table \ref{tab:tab2}) that outlines the complexity analysis for each of the models. In this complexity analysis, we have focused solely on the computation required per batched sample, without considering the choice of hyperparameters like the number of epochs or batch size. This provides a compact overview of the computational demands of each model on a per-sample basis.

\begin{table}[b]
  \centering
  \caption{Complexity of DNN, GRU, LSTM, RNN and CNN.  $t$ refers to the number of taps. $n_s$, $n_o$, $n_h$, and $d$ denotes input, output, hidden neuron, and depth of the DNN, respectively.}
  \label{tab:tab2}
\footnotesize   
\begin{tabular}{c|ccccc}
\toprule
Models & DNN & GRU & LSTM & RNN & CNN\\
\midrule
Train & $O(d-1)n^2)$  & $O((3n_h^2+6n_h)n_s)$ & $O(4n_h^2+7n_h)n_s$ & $O(n_h^2n_s)$ & $O(n_o)$\\
Inference  & $O(d-1)n^2)$  & $O((3n_h^2+6n_h)n_s)$ & $O(4n_h^2+7n_h)n_s$ & $O(n_h^2n_s)$ & $O(n_o)$\\
\bottomrule
\end{tabular}
\end{table}

\section{Advanced Sequential ML Methods}
\label{sec:Temporal Convolution Neural Network}
In Section \ref{sec:Traditional Sequential ML Methods}, we introduced traditional sequential models, such as RNN, LSTM, GRU, and CNN. The key question we aim to answer in this section is how to effectively incorporate the unique characteristics of time series data into the modeling process and leverage the temporal convolution model to mitigate channel distortion. Compared to other DL models like transformers and Fourier-based neural networks, convolutional models exhibit better generalization performance. Convolutional models are also more robust to changes in their parameter values when applied to new datasets, unlike the other models which require careful parameter initialization and hyperparameter tuning when used on new data \cite{liu2022scinet}.

This section starts with channel modeling, encompassing four distinct noise models. We derive the characteristics and capabilities required for the algorithm based on these models. Subsequently, we provide a detailed exposition of the architectures and fundamental assumptions underlying three models: Frequency-Calibrated Sampled-Interaction Neural Network (FC-SCINet) \cite{ref-fcscinet}), Light time-series (LightTS) \cite{zhang2022less} and DLinear \cite{dlinear}.\\

\subsection{Distortion Model:} The main limiting factor for the equalization task in a short-reach/PON system is ISI as a result of CD, sampling error (jitter), frequency shift (chirp), and Kerr-induced nonlinearity \cite{karanov2018end}. In this section, we will focus on the effects of CD, jitter, and chirp, as these are the dominant distortion mechanisms in short-reach PAM-based systems. The impact of Kerr-nonlinearities is limited in singe-channel PONs due to the relatively short fiber lengths and low optical powers involved.

\textbf{CD} in an optical communication system is caused by different phase velocities with respect to frequency. It fundamentally constitutes a linear transformation, and its mathematical representation involves a differential equation that considers spatial position and time, which can be presented as 
\begin{equation}
\label{equ: cd}
\frac{\partial A}{\partial z} = -j \frac{\beta_2}{2} \frac{\partial^2 A}{\partial t^2}
\end{equation}
where $A$ denotes the amplitude of the complex signal, $t$ denotes time, $z$ is the spatial position along the fiber, where the pulse pattern propagates \cite{karanov2018end}, and $\beta_2$ is the dispersion coefficient. Following the Fourier transformation, we have 
\color{black} 
\begin{equation} 
\label{equ: CD fourier}
D(z, w) = \exp(j \frac{\beta_2}{2} \omega^2 z)
\end{equation}
$w$ is the angular frequency. In the time domain, it is primarily manifested by significant attenuation in the high-frequency components and rapidly changing components.

\textbf{Jitter} is caused by fluctuations in sampling time. It presents itself as signal distortion, exhibiting a comparable impact to superimposed interference signals that adhere to the Gaussian distribution. Timing jitter can be described as  
\begin{equation} \label{eq:jitter}
\begin{split}
y_{*}(t)  &= y(t)\sum_{n=-\infty}^{+\infty}\delta(t-n*t_A-\tau) \\
 & = y(n*t_A+\tau) 
\end{split}
\end{equation}
where $\tau$ is the timing sampling error, where the correctly sampled value is $y(n*t_A)$. This sampling error can be quantitively estimated as follows:
\begin{equation} \label{eq:jitter estimation}
\vert y(n*t_A+\tau) - y(n*t_A) \vert \le M_1 \vert \tau \vert 
\end{equation}
where $M_1$ is the first moment of the band-limited spectrum of Fourier transformation of original signal $Y(f)$, can be simply written as:
\begin{equation} \label{eq:jitter estimation}
\vert \frac{\partial y(t)}{\partial t} \vert \le \sum_{-f_g}^{f_g} \vert 2\pi f \vert \vert Y(f)\vert = M_1  
\end{equation}
Jitter refers to high-frequency fluctuations in the amplitude of a signal. This high-frequency perturbation can have a significant impact on neural networks that rely on low-frequency signals. 

In conclusion, the error can be estimated as $|y(t_n) - y(n\tau_a)| \leq M_1|t_n - n\tau_a| = M_1|\tau_n|$.  The error $|e_n|$ is bounded by $M_1 \cdot |\tau_n|$ for a given $n$. The value of $e_n$ depends solely on $\tau_n$. Assuming that the timing error $\tau_n$ follows a statistical nature with $E\{\tau_n\} = 0$ and $E\{\tau_n^2\} = \sigma_{\tau}^2$, it follows that the amplitude errors $e_n$ are statistically independent. Consequently, the error variance is then given by $E \{ e_n^2 \} \leq M_1^2 \sigma_{\tau}^2$. For more details, please refer to \cite{kiencke}.

\textbf{Chirp} is a signal whose frequency varies with time. Mathematically, it can be described as follows, 
\begin{equation}
    s(t) = a(t) \cdot \exp[j(\omega_0 \cdot t + \theta(t))]
\end{equation}

The frequency spectrum of this waveform is obtained as 

\begin{equation}
S(\omega) = \int_{-\infty}^{\infty} a(t) \cdot \exp[j(\omega_0 t + \theta(t))] \cdot \exp(-j\omega t) \, dt
\end{equation}
Simplifying further:
\begin{equation}
S(\omega) = \int_{-\infty}^{\infty} a(t) \cdot \exp[j\{(\omega_0 - \omega)t + \theta(t)\}] \, dt
\end{equation}
In summary, all types of effects encountered in equalization issues, except for jitter, involve concurrent alterations in both the time and frequency domains. It is noteworthy that such changes are not statistically independent. Consequently, no single effect can be eliminated through straightforward nonlinear operations in a single domain. 

\begin{figure}[tb]
    \centering
    \includegraphics[width=0.75\linewidth]{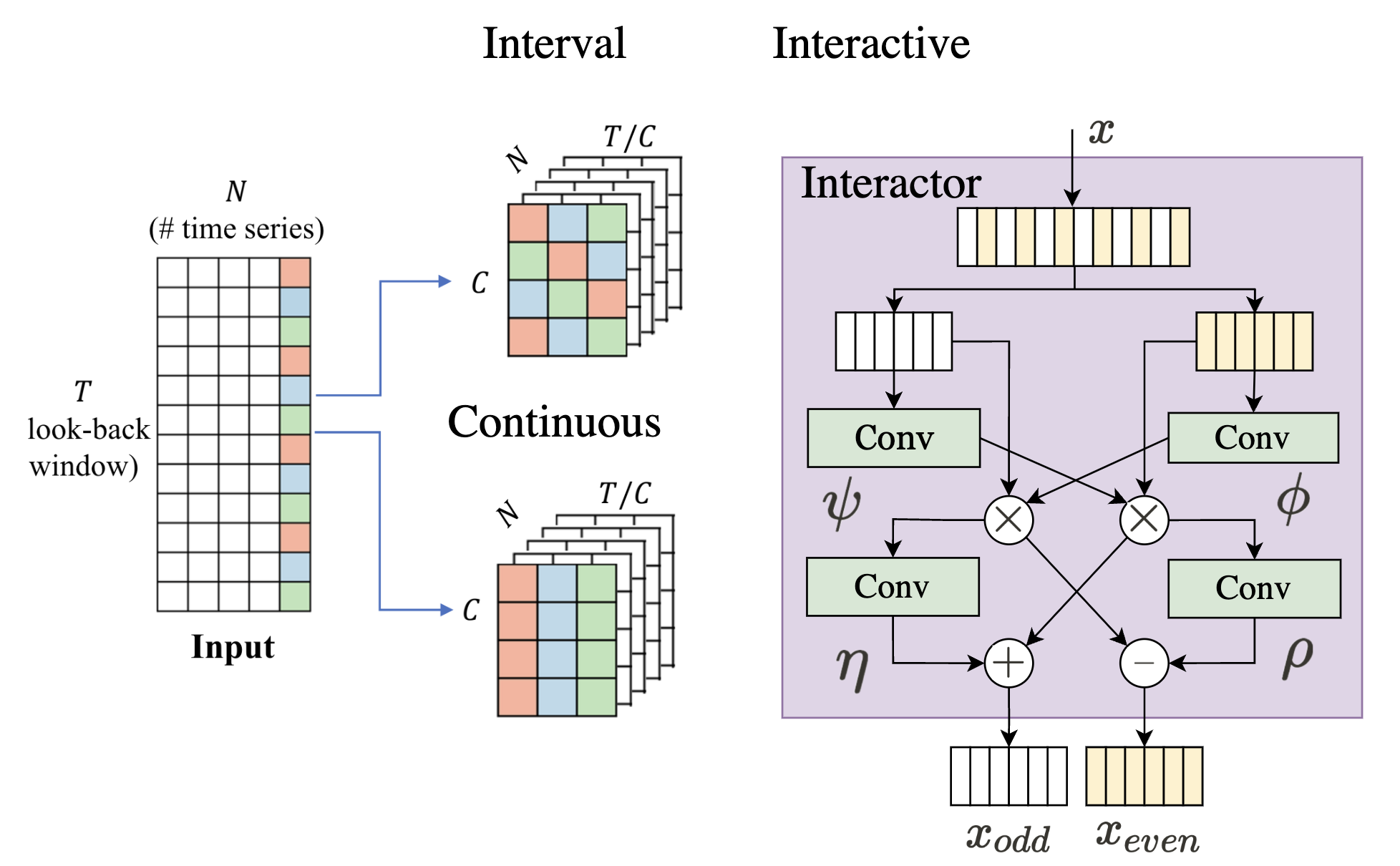}
    \caption{The overview of all sampling modules in temporal convolution networks is modified from \cite{ref-fcscinet, zhang2022less}, namely interval sampling and continuous sampling in LightTS \cite{zhang2022less}, and interactive sampling in \cite{ref-fcscinet, liu2022scinet}.}
    \label{fig:1}
\end{figure}

\subsection{Temporal Convolution Neural Network}
DL-based equalizers fundamentally capture domain-specific nonlinear disturbances by employing linear transformations and nonlinear activation functions. Within the temporal CNN, the core modules comprise interval, continuous, and interaction sampling modules, alongside convolutional neurons and linear layers, as depicted in Fig. \ref{fig:1}. Each of these modules offers practical flexibility for hardware implementation, due to their computational efficiency. Subsequently, we delve into three of the most efficient convolution-based sampling networks.

\textbf{FC-SCINet:} This novel approach introduces an improved series decomposition technique as a spectrum correction module. In conjunction with the interaction sampling module, it has proven to be a robust tool for mitigating CD and addressing various real-world channel effects \cite{ref-fcscinet}.

\textit{Decomp}: In the case of FC-SCINet, it utilizes a moving averaging filter with kernel size $w_1$ to extract low-frequency signals from the input. Additionally, high-frequency signals are obtained by calculating the residuals between the original and low-frequency signals. The final output signal is generated through a weighted linear combination of these two components, which is $\hat{x}$ defined as Eq.\eqref{eqn:equ2}.
\begin{align}
\label{eqn:equ2}
    \hat{x} = W_s^T x_s + W_f^T x_f
\end{align}
The complexity of this module is $O(k)$, where $k$ is the size of the kernel and is independent of the input sequence length. 

However, as demonstrated in the empirical study in \cite{ref-fcscinet}, the performance of FC-SCINet in mitigating CD remains strong. Moreover, the plug-and-play nature, low complexity, and interpretability of FC-SCINet make it highly flexible for seamless integration with various other algorithms. The DLinear architecture is another impressively low-complexity yet high-performance design.

\textit{SCIBlock:} The SCIBlock, is a key component of FC-SCINet, because it can iteratively decompose a signal into sub-sequences at various scales, while incorporating nonlinear transformations between adjacent layers. In contrast, the decomp block is restricted to enhancing fixed signal components and is limited to a single scale. From a mathematical perspective, the SCIBlock applies a hierarchical structure by systematically downsampling the input sequence into even-positioned and odd-positioned samples, denoted as $x_{even}$ and $x_{odd}$. Following the convolutional layer, the sub-sequences in adjacent layers are iteratively multiplied together utilizing exponential and multiplication operations, as shown in Eq. \eqref{eqn:equ3}  and  \eqref{eqn:equ4}.
\begin{align}
\label{eqn:equ3}
  x_{even}^{s} = x_{even}\odot \exp(\psi(x_{odd})), \quad  x_{odd}^{s} = x_{odd}\odot \exp(\phi(x_{even})) \\
\label{eqn:equ4}
  x_{odd}' = x^s_{even} + \exp(\eta(x^{s}_{odd})) \quad 
  x_{even}' = x^{s}_{odd} - \exp(\rho(x^{s}_{even})) 
\end{align}
Here, $\odot$ represents an element-wise product, and $\psi$, $\phi$, $\eta$, and $\rho$ are independent 1D convolutional layers. The intermediate outputs can be presented as $x_{even}^{s}$, $x_{odd}^{s}$, $x_{even}'$, and $x_{odd}'$. Upon completion of the processing, the resulting sub-sequences are then reassembled and aligned back to their original positions within the original signal. Ultimately, all the sub-series are concatenated based on their original index in the raw sequence, as illustrated in Fig. \ref{fig:1}.
To sum up, FC-SCINet is a framework capable of efficiently learning local-dependent patterns. Its distinctive feature lies in performing interactive learning on sub-sequences with odd-even positions after odd-even sampling, allowing for a larger receptive field under the premise of using the same convolutional kernel. \\ 

\textbf{DLinear:} As previously mentioned in FC-SCINet, while the concatenation of the decomp module may not offer optimal equalization, it has exhibited great performance in real-world datasets. Therefore, we will provide a brief introduction to this module: It first decomposes a raw data input into a low $ \mathbf{x_s} $ and high frequency $\mathbf{x_f}$ signal. $ \mathbf{x_s} $ is extracted by a moving average kernel. It is equivalent to filtering the signal using a $sinc$ function in the frequency domain. These two components are added in a linear combination form, expressed by $W_s$, $W_f$. The operation above is presented in Eq.\eqref{eqn:equ1}.
\begin{align}
\label{eqn:equ1}
  \mathbf{x_s} = \text{AvgPool}(\mathbf{x})  \quad 
  \mathbf{x_f} =  \mathbf{x} -  \mathbf{x_s}  \quad 
  \mathbf{x}'= W_s\mathbf{x} + W_f\mathbf{x_f}
\end{align}
By iterative decomposition with different kernel sizes, DLinear can be extended to a deeper network. The complexity is $O(kn_s)$, where $k$ is the number of the layer, and $n_s$ is the length of the model input. To simplify the complexity, the weight matrix $W$ could be replaced by the convolutional kernel. \\ 

\textbf{LightTS}: Both FC-SCINet and DLinear utilize only convolution and different sampling modules to capture the local and global dependencies. The LightTS architecture, detailed in \cite{zhang2022less}, employs a multi-layer perceptron (MLP) structure to enhance predictive abilities. 

\textit{Sampling:} In contrast to SCIBlock, which samples the raw sequence using odd and even indices, LightTS introduces two generic sampling strategies: Interval Sampling and Continuous Sampling, as shown in Fig. \ref{fig:1}. Interval sampling partitions time series data into non-overlapping sub-sequences based on fixed time intervals, as shown in Fig \ref{fig:1}. This approach helps identify periodic patterns or regularities within the data while minimizing information loss. On the other hand, continuous sampling divides sequences into corresponding sub-sequences, extracting data points continuously throughout the time series and preserving temporal continuity. This sampling method enables the capture of patterns within the period, ensuring a more comprehensive representation of the underlying dynamics. The subsequent section presents an MLP-based architecture to extract useful features from both the downsampled sub-sequences and continuously sampled sub-sequences.

\textit{Information Exchange Block (IEBlock)}: The IEBlock serves as the central module in LightTS, designed to effectively process the 2D matrix resulting from continuous sampling and interval sampling. This block comprises three essential components: 1) temporal projection, which identifies temporal features following continuous sampling; 2) channel projection, which captures inter-channel information following interval sampling; and 3) the exchange block, which integrates the information from the outputs mentioned above, facilitating information fusion. All of them utilize MLP as the nonlinear behavior learning module. The design of LightTS is notably concise, employing only two sampling modules and an MLP. On certain datasets, it surpasses the performance of FC-SCINet \cite{dlinear}.  \\ 

Compared to the models mentioned earlier, the FC-SCINet model requires less structural adaptation and pre-processing when applied in practical PON  applications. The FC-SCINet has been successful in recent PON-related work \cite{ref-fcscinet}.
Different from LSTM, which offers the advantage of ensuring information flow strictly from past to future, temporal CNN goes beyond this by modeling the global dependency between input and output, while also leveraging stacked causal convolution layers. Additionally, the FC-SCINet introduces interaction modeling, enabling the explicit capture of interactions between elements within a sequence, making it a more advanced alternative to LSTM. In addition to these benefits, CNNs and SCINet offer several advantages over LSTM: 
\begin{itemize}
    \item  CNNs can identify patterns regardless of their position within the input sequence. This property makes them well-suited for tasks where the position or timing of relevant features is not fixed, providing greater flexibility compared to LSTM.
    \item CNNs excel at capturing local patterns and extracting relevant features from the input sequence. This ability is particularly useful for tasks that require identifying and recognizing specific patterns or motifs within the data.
    \item Both CNNs and SCINet architectures typically have fewer parameters compared to LSTM models. This reduced parameter count can make training and inference more efficient, especially when working with limited computational resources or when dealing with large datasets.
\end{itemize}

\textbf{Recent Progress}: CNNs play a crucial role in current time series prediction research and applications. This is due to their high parameter efficiency, model stability, and strong theoretical foundation (Multiscale Decomposition). The complexity of these convolutional networks mostly depends on the number of layers and the size of the convolutional kernels. Nowadays, more advanced designs like dilated convolution and inception are often combined with other modules to create complex DL models, but they are rarely used on their own. Even so, temporal CNNs still have a distinct advantage in terms of the performance-to-complexity ratio. They are also straightforward to implement in hardware.

\subsection{Transformer-based Network}
\label{sec:Transformer}
\textit{Attention:} The Scaled Dot-Product Attention mechanism is the key component aiming to aggregate information across different parts of the input sequence. Each input vector is transformed into three distinct vectors: Queries ($Q$), Keys ($K$), and Values ($V$). The process involves calculating the dot products of the queries with all keys, scaling them by the square root of the dimension, and applying a softmax function to obtain weights on the values \cite{vaswani2017attention}. 

\begin{equation}
    \text{Attention}(Q, K, V) = \text{softmax}\left(\frac{QK^T}{\sqrt{d_k}}\right) V 
\end{equation}

The resulting matrix of outputs is obtained through a weighted sum of the values, where the weights are determined by the softmax-processed dot products of queries and keys. This attention module allows the model to focus on relevant parts of the input sequence, capturing local dependencies during the training process. A residual connection is applied around the two sub-layers, followed by layer normalization to maintain the information flow.

Transformers use multiple attention heads to look at the input sequence from different perspectives. This allows the model to simultaneously learn and consider various views of the input data. Equation \ref{eq_multihead_1} and \ref{eq_multihead_2} represent the functionality of the multi-head attention step. $head_i$ represents the single attention head. The final result of the multi-head attention is concatenating all the attention heads.

\begin{equation}
\label{eq_multihead_1}
\begin{aligned}    
\text{MultiHead}(Q,K,V)=\text{Concat}(\text{head}_1, \dots, \text{head}_h)W^o
\end{aligned}
\end{equation}
\begin{equation}
\label{eq_multihead_2}
\begin{aligned}    
\text{head}_i=\text{Attention}(QW_i^Q, KW_i^K, VW_i^V)
\end{aligned}
\end{equation}

While the standard ('vanilla') transformer model has shown great performance on time series data, the computational complexity of its attention mechanism makes it struggle to handle long sequences effectively. To overcome this limitation, researchers have developed various attention mechanism variants. An example is the Locality-Sensitive Hashing (LSH) Attention mechanism, which was introduced as part of the Reformer architecture \cite{kitaev2020reformer}. The LSH Attention mechanism utilizes specialized hash functions to transform queries and keys, thereby facilitating the categorization of similar items into shared hash buckets. Through sorting tokens based on their hash codes, items with similarities are grouped together, enabling the aggregation of relevant information. To enable parallel processing, the sorted sequence is divided into chunks. Subsequently, attention mechanisms are selectively applied to these chunks and their neighboring segments, allowing for focused examination of localized portions. The LSH Attention mechanism uses hash coding to greatly improve the computational efficiency of the transformer compared to the original version. This helps address the challenge of processing long sequences by reducing complexity without sacrificing performance.

\textbf{Decoder:} The decoder consists of several stacked sub-decoders. The ground truth follows a process similar to that of the encoder, being transformed into Query $Q$, Key $K$, and Value $V$ representations. The attention weights are calculated by comparing the Queries $Q$ from the decoder with every value $V$ in the encoder. This process is repeated in parallel across N sub-decoders, resulting in a final attention matrix. The attention matrix then undergoes a softmax operation, yielding probabilities for each value. In addition to the two sub-layers in each encoder layer, the decoder introduces a third sub-layer, performing multi-head attention over the encoder stack's output.

During the decoding process, the model is auto-regressive, using the previously generated outputs as additional input to generate the next output. Residual connections and layer normalization are used around each sub-layer. The self-attention sub-layer is modified to prevent positions from attending to future positions. This, along with the offset output embedding, ensures that the predictions for a position only depend on the known outputs at earlier positions in the sequence.

\textbf{Attention Variants:}
The purpose of the attention layer is to identify connections and dependencies among the various input embeddings. This allows the model to evaluate the importance of each element in relation to the others. The attention mechanism explicitly computes the relationships between different elements in the sequence, providing insights into how information flows through the model. However, except for the computational complexity issue, the mechanism in vanilla transformer \cite{vaswani2017attention} needs to be improved in terms of processing inter-dependencies and periodicity of signal data. The Autoformer model \cite{wu2021autoformer} introduces a new type of encoder that replaces the original encoder. This new encoder applies series decomposition and autocorrelation to detect dependencies between different parts of the input sequence, and then combines the representations of the sub-series. The series decomposition component divides the original signal into two distinct parts: the seasonal component, which captures short-term patterns, and the trend component, which captures long-term behavior. This partitioning allows for identifying and representing both the short-term and long-term characteristics present in the time series data. Additionally, the auto-correlation mechanism utilizes the fast Fourier transform (FFT) to compute correlations between the time series and its delayed version, providing insights into how the series relates to its past values at different time lags. The combination of series decomposition and autocorrelation effectively captures and represents the underlying trend and seasonality in the time series data.

\textbf{Recent Progress:} In this section, we provide a comprehensive overview of the vanilla transformer and its architecture, particularly within the domain of time series and traffic prediction. Over the years, substantial improvements have been made to enhance the transformer for accurate time series prediction. Notably, advancements have been achieved in reducing computational complexity while improving the effectiveness of the attention mechanism \cite{logsparse2019, informer2021}. However, recent research has introduced compact models based on multi-scale transformation \cite{scaleformer2023}, which surprisingly outperforms benchmark-designed models. This new development has sparked an important debate on the fundamental structure of sequence models. In the following sections, we summarize and explore this particular model in depth, providing insights into its implications. For the latest work, please refer to Table 3 (table \ref{tab:tab4}).

\subsection{Fourier Convolution Neural Network}
\label{sec:Fourier Convolution Neural Network}
In the previous section, we introduced models based on convolutional kernels and subsampling as fundamental modules. Their core principle involves decomposing signals into different scales in the time domain and subsequently applying nonlinear transformations to learn salient features. However, for the majority of real-world signals, transforming them into the Fourier domain is often more efficient. This efficiency is attributed to the following factors: 1) The majority of real-world signals are bandpass or lowpass, and in the Fourier domain, their dynamic range decreases from $n$ to $exp(-n)$; 2) The Fourier transformation is a bijective (one-to-one) transformation, which ensures energy conservation and controllable error in both the forward and inverse transformations; 3) The computational complexity of existing (fast) Fourier transform algorithms, after improvements, is $O(nlog(n))$, making it convenient for hardware implementation. \

In this section, we introduce TimesNet \cite{TimesNet} which utilizes a frequency-attention mechanism and FreTS \cite{yi2023frequency} which explicitly performs non-linear transformations in the frequency domain. For extensive models please refer to Table \ref{tab:tab4}. The fundamental concept of TimesNet involves transforming the initial signal into $k$ distinct 2D tensors instead of directly processing the original sequence. This approach empowers the model to effectively capture both intra-periodic and inter-periodic variations within these fixed windows. A variant of this model has been recently reported in \cite{shao2024advanced}

\textbf{TimesNet:} An attention mechanism based on spectral amplitude is employed to determine the significance of signal segments at various frequencies. Simultaneously, across different temporal resolution scales, a shared convolution module is utilized to reconstruct nonlinear distortions introduced by the channel. It does not explicitly perform nonlinear transformations in the frequency domain; instead, it combines reconstructed signals at different window lengths through a linear combination. The FConvNet primarily comprises four key blocks: Component Detection, Alignment, ConvNet, and Reconstruction.

\textit{Component Detection}: The identification of the k most crucial frequencies is based on the amplitude of the Fourier coefficients. Then, using only the selected components within the k frequency range, the signal is sampled using a continuous sampling method, and these sub-series are arranged into a two-dimensional tensor.

\textit{Alignment}: The aligned sub-series are then fed into a convolution-based module, specifically an Inception network, to mitigate distortion caused by channel effects. The Fourier coefficients pass through a softmax function to generate attention weights, which are then multiplied by the output of each convolution module to produce the final output.

\textit{Fourier Attention}: The Fourier transformation is a global operation, meaning that any changes in the signal's amplitude will cause periodic oscillations throughout the entire signal. Significant variations in the amplitude of the primary components lead to substantial fluctuations. TimesNet leverages this characteristic by using the Fourier spectrum of the nonlinearly transformed signal to determine the attention value for each component.

\textit{Reconstruction}: Finally, employing a residual form, we obtain the reconstructed individual sub-components multiplied by their respective attention values, denoted as $y'$, and add them to the input signal $x$ to yield $y$.

\textbf{FreTS: } Time-domain-based processing is limited by information bottlenecks, as the local characteristics vary. FreTS explicitly uses frequency-domain features in its model architecture to directly mitigate distortion without manipulating the time-domain. FreTS is essentially an MLP-based network that is able to effectively learn patterns of time series data in the frequency domain. As presented in \cite{yi2023frequency}, FreTS consists of two learners: a Frequency Channel Learner and a Frequency Temporal Learner. In the equalization problem, there is no actual channel dimension, but rather a stack of independent experiments. Therefore, FreTS only introduces a frequency domain MLP.

\textit{Frequency MLP}: The frequency temporal learner aims to capture temporal patterns in the frequency domain. Specifically, for a complex number input $\mathcal{H} \in \mathbb{C}^{m \times d}$,  the MLP aims to optimize the weight matrix $\mathcal{W} \in \mathbb{C}^{d \times d}$ and bias $\mathcal{B} \in \mathbb{C}^{d}$ so that the final output $\mathcal{Y} \in \mathbb{C}^{m \times d}$ could approximately reconstruct the ground truth.
\begin{align}
\mathcal{Y}_{\ell} & = \sigma(\mathcal{Y}_{\ell-1}\mathcal{W}_{\ell} + \mathcal{B}_{\ell})  \\
\mathcal{Y}_0 & = \mathcal{H}
\end{align}
The MLP in the frequency domain is equivalent to global convolutions in the time domain as detailed in \cite{yi2023frequency}. An increasing number of studies have demonstrated the feasibility of DL models operating in the frequency domain. Simultaneously, the corresponding computational complexity of frequency-domain processing has decreased from $O(n)$ to $O(n log n)$ due to the reduction in the signal's dynamic range. However, the advantages and disadvantages of networks in both the frequency and time domains remain inadequately explored. Due to space limitations, we offer a detailed categorization, along with corresponding references and keywords, in Table \ref{tab:tab4} for researchers with specific interests.

\begin{table}[H]
\centering
\tiny
\caption{Benchmark Models}
\label{tab:tab4}
\footnotesize    
\begin{tabular}{>{\centering}p{1.5cm}|>{\centering\arraybackslash}p{2.3cm}|>
{\arraybackslash}p{3cm}|>
{\arraybackslash}p{5.2cm}}

\toprule
\multicolumn{2}{c}{Models} & Efficient Techniques & Literature \\
\midrule
\multirow{13}{*}{Transformer} 
& \multirow{7}{*}{Attention} 
& Sparsity inductive bias & \cite{logsparse2019} LogTrans leverages convolutional self-attention for improved accuracy with $O(L(\log L)^2)$ lower memory costs. \\
\cline{3-4}
& & Low-rank property & \cite{informer2021} Informer selects dominant queries based on queries and key similarities. \\
\cline{3-4} 
& & Learned Rotate attention (LRA) & \cite{Quatformerchen2022} Quatformer introduces learnable period and phase information to depict intricate periodical patterns. \\
\cline{3-4}
& & Hierarchical pyramidal attention & \cite{Pyraformer2022} Pyraformer proposed one hierarchial attention mechanism with a binary tree following the path with linear time and memory complexity\\
\cline{3-4}
& & Frequency Attention & \cite{Fedformer2022} FEDformer: proposed the attention operation with Fourier transform and wavelet transform. \\
\cline{3-4}
&& Correlation Attention & \cite{wu2021autoformer}Autoformer: the Auto-Correlation mechanism to capture sub-series similarity based on auto-correlation and seires decomposition\\
\cline{3-4}
& & Cross-dimension dependency & \cite{zhang2023crossformer} Crossformer utilizes multiple attention matrices to capture cross-dimension dependency \\
\cline{2-4}
& \multirow{3}{*}{Architecture} & Triangular patch attention & \cite{Cirstea2022TriformerTV} Triformer features a triangular, variable-specific patch attention with a lightweight and linear complexity  \\
\cline{3-4}
& & Multi-scale framework & \cite{scaleformer2023}Scaleformer iteratively refines a forecasted time series at multiple scales with shared weights\\
\cline{3-4}
& \multirow{2}{*}{Positional Encoding} & Vallina Position  &  \cite{vaswani2017attention} $\cos$ and $\sin$ functions with a sampling rate-relevant period.\\
\cline{3-4}
& & Relative Positional Encoding & \cite{Zerveas2021} Introduces an embedding layer that learns embedding vectors for each position index. \\
\cline{3-4}
& & Model-based learned & \cite{logsparse2019} LogSparse utilize one LSTM to learn relative position between series tokens\\
\midrule 
\multirow{5}{*}{Fourier-NN} & \multirow{2}{*}{Time Domain}  & Series Decomposition & \cite{dlinear} DLinear performs one linear series decomposition with multiple layers \\
\cline{3-4}
& & Frequency Attention & \cite{TimesNet} TimesNet proposes the attention mechanism related to the amplitude of the signal \\
\cline{2-4}
& \multirow{2}{*}{Frequency Domain}  & Frequency MLP & \cite{yi2023frequency}  FreqMLP performs MLP in frequency domain by leveraging the global view and energy compaction characteristic \\
\midrule
\multirow{11}{*}{TConv-NN} & \multirow{4}{*}{Sampling}  & Continous & \cite{zhang2022less} and \cite{TimesNet} both utilize continous sampling to split original signal into windowed subseries similar to short time transformation \\
\cline{3-4}
& & Interval & \cite{zhang2022less} Interval sampling with fixed step to extract periodic feature \\ 
\cline{3-4}
& & Even-Odd/Multiscale & \cite{ref-fcscinet} proposes one iterative multiscale framework where even and odd series are interacted between layers \\ 
\cline{3-4}
& & Frequency Continous & \cite{dlinear} leverages series decomposition module in a iterative manner to decompose signal in frequency domain with $sinc$ function. \\
\cline{3-4}
& & Negative sampling & \cite{TCN2019} custom loss function is employed in an unsupervised manner, wherein distant or non-stationary subseries maximize the loss, while similar subseries minimize the loss. \\
\cline{2-4}
& \multirow{4}{*}{Feature Module}  & MLP & \cite{zhang2022less} applies an MLP-based structure to both interval sampling and continuous sampling for extracting trend and detail information. \\
\cline{3-4}
&  & Dilated convolutions & \cite{Wu2019GraphWF} leverages stacked dilated casual convolutions to handle spatial-temporal graph data with long-range temporal sequences \\

\bottomrule
\end{tabular}
\end{table}

\section{Model Compression}
\label{sec:Model Compression}
In recent years, the proliferation of large-scale ML models has significantly advanced state-of-the-art technology across various domains, ranging from natural language processing to computer vision. The surge in model complexity, often characterized by sophisticated architectures and many parameters, has driven the need for efficient hardware implementations to harness their full potential. The advent of single graphics processing units (GPUs) as a critical computational resource has been pivotal, offering a parallelized architecture suitable for accelerating the training and inference processes \cite{lew2019analyzing}. The significance of deploying large ML models on a single GPU lies in optimizing computational efficiency and reducing latency. Single GPU implementations facilitate parallel processing, enabling the simultaneous execution of multiple tasks and handling extensive model parameters. This enhances the speed of model training and facilitates real-time inference, a critical requirement in applications such as autonomous systems and edge computing. However, the hardware implementation of large ML models on a single GPU is challenging. The complexity of these models often exceeds the computational capacity and memory constraints of a single GPU, necessitating innovative solutions for efficient utilization \cite{marculescu2018hardware}. Techniques such as model pruning, quantization, vector quantization, and knowledge distillation have emerged as strategies to mitigate these challenges, ensuring that even formidable models can be accommodated within the limitations of a single GPU without compromising performance. The authors in \cite{marculescu2018hardware} examine how to use a single GPU effectively for implementing large ML models. They discuss methods that balance complexity and computational efficiency to maximize hardware utilization \cite{marculescu2018hardware}.

In addition, conducting a comprehensive performance-versus-complexity analysis is necessary to evaluate the suitability of various ANNs in short-reach optical communication systems. DL models, including CNNs, RNNs, and LSTMs, find applications in critical tasks such as equalization, fault detection, subcarrier allocation, nonlinearity compensation, and bandwidth request and allocation. The complexity of these models is a significant factor affecting their feasibility. For instance, CNNs may introduce convolutional and pooling layers, increasing model complexity. Similarly, RNNs and LSTMs, designed for sequential data, introduce recurrent connections that enhance their ability to capture temporal dependencies and contribute to increased complexity \cite{ff9483687}. Analyzing the neural network architectures in detail, including their depth, the number of parameters, and computational demands, is crucial for understanding the trade-offs between performance and complexity. DL models often exhibit enhanced capabilities in capturing complex patterns and relationships in optical communication data. Still, their complexity may pose challenges regarding training time, computational resources, and practical implementation \cite{app9152975}. A thorough examination of these complexities is essential for identifying optimal models, such as choosing between a CNN for image-based tasks or an LSTM for sequential data, that balance high performance and manageable complexity, facilitating their efficient integration into short-reach optical communication systems \cite{ff9483687}. Four prominent types, namely the feed-forward neural networks (FFNN), the radial basis function neural networks (RBF-NN), the auto-regressive RNN (AR-RNN), and layer-RNN (L-RNN), offer distinct trade-offs in complexity and performance. Among nonlinear neural network-based equalizers with equivalent numbers of inputs and hidden neurons, FFNN-based equalizers have the lowest computational complexity, however AR-RNN demonstrate superior transmission performance in 50 Gb/s PAM4 systems \cite{Xu:19}.

\textbf{Distillation model}: Knowledge distillation, a model compression technique, transfers knowledge from complex, large-scale models or groups to more compact, feasible models suitable for real-world applications. Pioneered by Bucilua et al. in 2006 \cite{knowledge2006}, knowledge distillation primarily operates on neural network architectures characterized by multifaceted structures comprising multiple layers and parameters. Knowledge distillation has been recently considered an important technique for practical DL applications such as speech recognition, image recognition, and natural language processing \cite{gou2021knowledge}. Deploying large deep neural network models can be especially challenging for edge devices, which are limited in memory and computational power. To address this challenge, an innovative model compression method was developed in \cite{gou2021knowledge}, allowing transferring knowledge from larger, more complex models to train smaller, more efficient models without significant performance loss. This process, where a smaller model learns from a larger one, was formalized into the "Knowledge Distillation" framework by Hinton et al. \cite{hinton2015distilling}. This framework has become crucial for deploying the essential knowledge from sophisticated, large-scale models on computationally constrained edge devices.

Optimizing DL models through knowledge distillation shows great potential for advancing short-range optical communication systems. RNNs have been particularly effective at addressing nonlinear distortions \cite{ff9483687, deligiannidis2021performance}. However, the feedback loop inherent in RNN structures makes it difficult to parallelize them, preventing their use in low-complexity hardware designed for high-speed processing in optical networks \cite{chang2017hardware}. Using knowledge distillation is a promising approach to enable parallelization of RNNs \cite{ff9483687, srivallapanondh2023knowledge}. This application of knowledge distillation is set to revolutionize the implementation of RNNs, ensuring compatibility with low-complexity hardware and meeting the stringent processing requirements of high-speed optical networks \cite{robey2021parallel}.

Beyond just RNNs, knowledge distillation can be applied to many different ML models important for optical communications, such as CNNs, LSTMs, FFNNs, RBF-NNs, AR-RNNs, and L-RNNs \cite{srivallapanondh2023knowledge}. These models each have their own challenges regarding complexity, adaptability, and real-time implementation. For example, using knowledge distillation in LSTMs for optical communication systems, can reduce model complexity without losing the ability to handle time-dependent patterns \cite{srivallapanondh2023knowledge}.

Another promising application of knowledge distillation is when facing challenges with limited time-series data. As "big data" impacts various fields, the scarcity of target events or high data acquisition costs can hinder ML in certain scenarios. A proposed method uses "privileged information" from partial time-series data during training to enhance long-term predictions for small datasets. Applied to optical communications, this distillation approach offers a solution to data constraints, demonstrating effectiveness on both synthetic and real-world data \cite{dd8851687}.

\textbf{Vector quantization}: Vector quantization (VQ) is a model compression technique targeting large-scale ML models. VQ represents complex data with a small set of prototype vectors, significantly cutting the computational load during inference. This makes VQ useful for applications that require balanced model efficiency and performance, such as when resources are limited. The VQ process involves partitioning the input space into regions, each with a representative prototype vector. During encoding, input vectors are assigned to the nearest prototype, quantizing the data. In the decoding or reconstruction phase, these prototype vectors are used to rebuild the original data.

The effectiveness of VQ relies on carefully selecting and updating the prototype vectors. The goal is to optimize the prototypes so they can effectively capture the essential information in the dataset \cite{gray1984vector}. By clustering and quantizing input vectors into a representative codebook, VQ enables encoding information in a more compact form. This is particularly beneficial in scenarios with limited data availability or high computational demands \cite{pourghasemi2020application}. For example, VQ can be useful when applying knowledge distillation to RNNs. RNNs face challenges with parallelization due to their feedback loop structure. Using VQ in the distillation process for RNNs can help address the parallelization issue. VQ can represent the essential information from the RNN using a smaller set of prototype vectors. \cite{Xu:19, wang2019vector, rasul2022vq}. This compression not only aids in overcoming hardware complexity but also contributes to faster processing in high-speed optical networks.

VQ uses an iterative process to improve the prototype vectors and enhance their ability to represent the data. Commonly, algorithms like k-means clustering are used for this. The prototypes are adjusted to minimize the difference between the original data and the quantized representation. This iterative refinement allows VQ to adapt to the patterns and structures in the data. This optimizes the compression capabilities of VQ while still preserving the critical information needed for training tasks \cite{gray1984vector}.

Finally, VQ can be beneficial in optimizing other DL models, such as CNNs or LSTMs, by efficiently capturing essential features with a minimal set of representative vectors \cite{ozair2021vector}. Exploring the use of VQ together with these models provides a promising way to improve the performance and scalability of ML applications in short-reach optical communication systems.


\section{Conclusion}
\label{sec:Conclusion}
In this survey, we have undertaken a comprehensive examination of powerful machine learning models that exhibit the potential to achieve robust equalization in cost-sensitive short-reach optical systems, with a particular focus on PONs. Our objective has been to explore these models' capacity to operate efficiently and deliver effective computational performance. For the first time, we have classified the current models into three distinct types and conducted an extensive analysis of their core concepts, highlighting their differences, similarities, and the underlying insights they provide. Additionally, we have presented a simplified complexity analysis considering various input sizes. In the final stages of our survey, we have also investigated the potential of machine learning solutions in addressing the challenges associated with hardware implementation and complexity. We firmly believe that this survey will serve as a valuable resource, inspiring future research endeavors to develop efficient models explicitly tailored for short-reach and PON systems.


\begin{adjustwidth}{-\extralength}{0cm}

\reftitle{References}

\PublishersNote{}
\end{adjustwidth}
\end{document}